\documentclass[aps,prl,unsortedaddress,
               showpacs,footinbib,
               amsfonts,amssymb,amsmath,nobibnotes,
               twocolumn, reprint,
              ]{revtex4-1}
\usepackage[utf8]{inputenc}
\usepackage{mathrsfs}
\usepackage{bm}  
\usepackage{graphicx}
\usepackage{color}
\usepackage{braket}
\usepackage{placeins}
\usepackage{pbox}
\usepackage{lipsum}

\usepackage[breaklinks]{hyperref}

\usepackage[hyphenbreaks]{breakurl}

\hypersetup{
  colorlinks=true,  
  linkcolor=blue,   
  citecolor=blue,   
  filecolor=black,  
  urlcolor=blue     
  }

\newcommand{\comment}[1]{}

\definecolor{lightgray}{gray}{0.8}
\definecolor{midgray}{gray}{0.5}
\definecolor{darkgray}{gray}{0.3}
\definecolor{darkergray}{gray}{0.2}
\definecolor{darkestgray}{gray}{0.1}

\definecolor{MidPaleAqua}{RGB}{63,128,155}
\definecolor{EmpireGreen}{RGB}{39,81,42}
\definecolor{RoyalBlue}{RGB}{29, 56,126}
\definecolor{VertDeGris}{RGB}{59,117,71}
\definecolor{Tomette}{RGB}{155,65,49}
\definecolor{MidAqua}{RGB}{50,119,178}

\let\oldlipsum\lipsum                                                            
\renewcommand{\lipsum}[1][1]{{\color{lightgray}{\oldlipsum[#1]}}}

\newcommand{\mv}[1]{\mathbf{#1}}

\newcommand{\ieig}{\vk n}

\newcommand{\eig}{\varepsilon_{\ieig}}

\newcommand{\icell}{l}
\newcommand{\iat}{\kappa}
\newcommand{\icart}{j}

\newcommand{\jat}{\iat'}
\newcommand{\jcart}{\icart'}

\renewcommand{\vr}{\mv{r}}
\newcommand{\vR}{\mv{R}}
\newcommand{\vq}{\mv{q}}

\newcommand{\vk}{\mv{k}}

\newcommand{\ipho}{\nu}

\newcommand{\etal}{{\it et al. }}

\newcommand{\berkeley}{%
  Department of Physics, University of California at Berkeley,
  California 94720, USA
  and Materials Sciences Division, Lawrence Berkeley National Laboratory,
  Berkeley, California 94720, USA
    }

\newcommand{\berkeleymail}{antonius@lbl.gov} 

\begin{document}

\title{Temperature-induced topological phase transitions:\\
       promoted vs. suppressed non-trivial topology}
\author{Gabriel Antonius} \email{\berkeleymail} 
\author{Steven G. Louie} 

\affiliation{\berkeley}

\begin{abstract}

We determine the topological phase diagram
  of BiTl(S$_{1-\delta}$Se$_{\delta}$)$_2$
  as a function of doping and temperature from first-principles calculations.
Due to electron\textendash phonon interaction,
  the bands are renormalized at finite temperature,
  allowing for a transition between the trivial ($Z_2=0$)
  and non-trivial ($Z_2=1$) topological phase.
We find two distinct regions of the phase diagram with non-trivial topology.
In BiTlS$_2$, the phonons promote the crystal to the topological phase
  at high temperature, while in BiTlSe$_2$, the topological phase exists
  only at low temperature.
This behaviour is explained by the symmetry of the phonon coupling potential,
  whereby the even phonon modes (whose potential is even under inversion)
  promote the topological phase
  and the odd phonon modes promote the trivial phase.

\end{abstract}

\maketitle

Recent studies on three-dimensional topological insulators
  have identified several materials with tunable topological phases
  \cite{hasan_topological_2015}.
Upon varying experimental parameters,
  these materials undergo a phase transition
  between a trivial and a topological insulator state.
Such transition may occur
  as a function of impurity doping
  \cite{hsieh_topological_2008,Xu2011a,dziawa_topological_2012,
    xu_hedgehog_2012},
  pressure \cite{xi_signatures_2013,Bahramy2012,li_robust_2016},
  or temperature \cite{dziawa_topological_2012, reijnders_optical_2014,
    wojek_band_2014, wojek_direct_2015, zhang_electronic_2016}.
The effect of temperature becomes especially important
  for devices that are expected to operate under varying conditions
  \cite{zhu_topological_2013}.
It is thus desirable to be able to predict the topological
  phase diagrams of these materials and their physical origin.

Electron\textendash phonon inter\-actions underly
  the temp\-er\-ature-induced topological phase transition.
As more phonons are being thermally activated,
  the electronic band energies may shift and close the band gap
  until a band inversion occurs at some critical temperature.
This process was first described in 2D and 3D topological insulators
  from model hamiltonians  
 \cite{garate_phononinduced_2013,saha_phonon-induced_2014,li_conductivity_2013,
       lizhou_electronphonon_2015,yoshida_restoration_2016}.
First-principles calcualtions later confirmed that lattice deformation
  due to phonons could flip the $Z_2$ invariant
    \cite{kim_topological_2015}\footnote{
      B. Monserrat and D. Vanderbilt also reported
      first-principle topological phase diagram calculations
      in the Bi$_2$Se$_3$ family compounds,
      as a function of pressure and temperature
      \cite{monserrat_temperature_2016}.
      }.

  One remarkable prediction from Garate \etal
    \cite{garate_phononinduced_2013,saha_phonon-induced_2014}
    was that electron\textendash phonon coupling could induce
    a trivial to topological phase transition with increasing temperature.
  The requirement for this scenario to happen is
    a negative temperature coefficients for the band edge states
    in the trivial phase,
    which promotes a band inversion at high temperature
    and stabilizes the topological phase. 
  They proposed that such phenomenon could be seen
    in BiTl(S$_{1-\delta}$Se$_{\delta}$)$_2$,
    due to the presence of light atoms
    and the tunability of the band gap with doping.
  While no temperature-dependent measurements have been reported in this
    particular material, those performed in Pb$_{1-\delta}$Sn$_{\delta}$Se
    indicate the opposite trend\textemdash that the system goes back
    from a topological to a trivial phase at higher temperature
    \cite{dziawa_topological_2012, wojek_band_2014, wojek_direct_2015}.

  In this Letter, we compute from first-principles the topological phase diagram
    of BiTl(S$_{1-\delta}$Se$_{\delta}$)$_2$.
  The electron\textendash phonon coupling and the temperature dependence
    of the electronic band energies is obtained from
    density functional perturbation theory (DFPT)
    \cite{Baroni2001,Baroni1987,Gonze1997,Gonze1997a},
    and we simulate doping with a linear mixing scheme.
  We show that the electron\textendash phonon interaction
    causes a topological transition in the studied material,
    and indeed promotes the topological phase in BiTlS$_2$.
  However, this feature depends on the doping content.
  The opposite trend is predicted in BiTlSe$_2$, that is,
    the topological phase is suppressed at high temperature.

  \subsection*{Theory and methodology}

  As a result of the electron\textendash phonon coupling,
    the electronic energies acquire a temperature dependence
    given by
    \begin{equation} \label{tdepeig}
      \eig(T ) = \eig^0 + \sum_{\ipho} \int \frac{d\vq}{\Omega_{BZ}}
                   \Sigma_{\vk n}^{ep}(\vq, \ipho)
                   \big[ \ n_{\vq \ipho} (T ) + \tfrac{1}{2} \ \big]
    ,
    \end{equation}
  where $\vk$ and $n$ label the wavevector and band index of an electronic state,
    $\vq$ and $\ipho$ label the wavevector and branch index of a phonon mode,
    and $\Omega_{BZ}$ is the volume of the Brillouin zone.
  In this expression, the electron\textendash phonon coupling self energy
    has been decomposed
    into the individual phonon modes' contributions.
  As we made use of the adiabatic approximation, 
    all the temperature dependence of the electronic energies
    comes from the Bose-Einstein distribution of the occupations
    of the phonon modes $n_{\vq \ipho} (T )$,
    and the $\tfrac{1}{2}$ factor in Eq.\eqref{tdepeig}
    accounts for the zero-point renormalization.
  In the static theory of Allen, Heine and Cardona
    \cite{Allen1976,Allen1981,Allen1983}, 
    the contribution of a phonon mode to the self energy is
    \begin{equation} \label{sigma}
      \Sigma_{\vk n}^{ep}(\vq, \ipho) = \sum_{n'}
          \frac{\vert g_{\vk n n'}(\vq,\ipho)\vert^2}
               {\varepsilon_{\vk n} - \varepsilon_{\vk+\vq n'} + i\eta}
        - \frac{\vert g^{DW}_{\vk n n'}(\vq, \ipho)\vert^2}
               {\varepsilon_{\vk n} - \varepsilon_{\vk n'} + i\eta}
    ,
    \end{equation}
  where $g_{\vk n n'}(\vq,\ipho)$ are the electron\textendash phonon coupling
    matrix elements and $\eta$ is a small positive real number.
  The first and second terms of Eq.\eqref{sigma} are called the Fan
    and Debye-Waller term respectively.

  In most semiconductors and insulators, the self energy is positive for
    the last occupied band (i.e. reducing the hole energy)
    and negative for the first unoccupied band
    (i.e. reducing the quasi-electron energy).
  The band gap therefore closes with increasing temperature.
  The rationale behind this behavior is that,
    for a large band gap semiconductor, the top of the valence band
    would be repelled by the nearby occupied states with lower energies,
    while the bottom of the conduction band would be repelled by the
    nearby unoccupied states with higher energies.
  In the case of a topological insulator,
    the small band gap allows for a phonon-mediated interaction
    between the occupied and the unoccupied bands
    (since they are close in energy),
    and one has to give more consideration to anticipate
    the sign of the self-energy corrections.

  Depending on the sign of the electron\textendash phonon coupling
    induced self energy in Eq.\eqref{sigma},
    two possible scenario can occur with profound implications
    on the stability of the topological phase.
  In one case, the self energy would cause the band gap of a trivial insulator
    to close with increasing temperature, until a band inversion occurs,
    and the system reaches a topological phase at some critical temperature.
  At higher temperature, the inverted gap would further increase,
    thus stabilizing the topological phase.
  In the converse scenario, a system that is a topological insulator
    at low temperature could have its band gap shrink at higher temperature
    until the bands are re-inverted and the system reaches a trivial phase.
  We show here that which one of these scenario occurs depends on the details
    of the system under consideration; it could even be reversed as the
    doping changes.

  The method to compute phonon-related properties
    using DFPT is well established \cite{giustino_electron-phonon_2016}.
  Besides providing the thermodynamical properties of solids,
    it has been successfully applied to the temperature dependence
    of electronic band structures
    \cite{Marini2008,Giustino2010,Antonius2014,Ponce2014,
          antonius_dynamical_2015,monserrat_correlation_2016}.
  In this work, we employ a linearized scheme to interpolate
    the phonon-related quantities at intermediate doping
    between two reference crystals structures.

  The crux of the DFPT method for the electron-phonon coupling
    is the self-consistent calculation of the
    potential created by moving the atoms of the crystal
    in a periodic but non-commensurate unit amplitude displacement
    with wavevector $\vq$:
    \begin{equation}
      V_{\iat\icart}(\vq, \vr) = \sum_{\icell} e^{i \vq \cdot \vR_{\icell}}
        \frac{\partial V^{\text{\tiny SCF}}(\vr)}
        {\partial \tau_{\icell\iat\icart}}
    ,
    \end{equation}
    where $\icell$ labels a unit cell with lattice vector $\vR_\icell$,
    $\iat$ labels an atom within the unit cell,
    $\icart$ labels a Cartesian direction,
    and $\bm{\tau}$ is the position of an atom.
  From this periodic perturbation potential and the corresponding
    perturbed density, one evaluates the dynamical matrix,
    defined as the second-order derivative of the total energy with respect
    to unit displacements of two atoms.
  Its Fourier transform at wavevector $\vq$ is given by
    \begin{equation} \label{dynmat}
      \Phi^{\icart\jcart}_{\iat\jat}(\vq) =
        \sum_{\icell} e^{i \vq \cdot \vR_{\icell}}
        \frac{\partial^2 E}
        {\partial \tau_{\icell\iat\icart} \partial \tau_{0\jat\jcart}}
    .
    \end{equation}
  The equation for the phonon modes with energies $\omega_{\vq\ipho}$
    and polarization vectors $\xi^{\ipho}_{\iat\icart}$
    is then
    \begin{equation} \label{dyneq}
      M_\iat \ \omega_{\vq\ipho}^2 \ \xi^{\ipho}_{\iat\icart}(\vq) =
        \sum_{\jat\jcart}
        \Phi^{\icart\jcart}_{\iat\jat}(\vq) \ \xi^{\ipho}_{\jat\jcart}(\vq)
    ,
    \end{equation}
    where $M_\iat$ is the atomic mass.

  Once the phonon modes and the perturbation potential are known,
    the electron\textendash phonon self energy can be constructed.  
  Defining an electron\textendash phonon squared coupling matrix as
    \begin{align} \label{Omega}
      \Omega^{\iat\icart,\jat\jcart}_{\vk n n'}(\vq) = & \nonumber\\
          \bra{\vk n} V^*_{\iat\icart} & (\vq,\vr) \ket{\vk+\vq n'}
          \bra{\vk+\vq n'} V_{\jat\jcart}(\vq,\vr)\ket{\vk n}
    ,
    \end{align}
  we may write the squared electron\textendash phonon coupling matrix elements
    as
    \begin{equation}
      \vert g_{\vk n n'}(\vq,\ipho)\vert^2 =
        \frac{1}{\omega_{\vq\ipho}}
        \sum_{\iat,\jat} \sum_{\icart,\jcart}
        \Omega^{\iat\icart,\jat\jcart}_{\vk n n'}(\vq)
        \big[
            \xi^{\ipho}_{\iat\icart}(\vq) \  \xi^{\ipho *}_{\jat\jcart}(\vq)
        \big]
    \end{equation}
  and their Debye-Waller counterpart as
    \begin{align}
      \vert g^{DW}_{\vk n n'}(\vq, \ipho)\vert^2 = & 
        \frac{1}{2\omega_{\vq\ipho}}
        \sum_{\iat,\jat} \sum_{\icart,\jcart}
        \Omega^{\iat\icart,\jat\jcart}_{\vk n n'}(0) \nonumber\\
        & \times \big[
          \xi^{\ipho}_{\iat\icart}(\vq) \  \xi^{\ipho *}_{\iat\jcart}(\vq) +
          \xi^{\ipho}_{\jat\icart}(\vq) \  \xi^{\ipho *}_{\jat\jcart}(\vq)
        \big]
    .
    \end{align}

  We perform electronic structure and DFPT calculations
    on the reference systems BiTlS$_2$ and BiTlSe$_2$,
    and we use a linear mixing scheme as the simplest model
    for an intermediate doping. 
  To simulate a doping $\delta$ resulting in
    the stoechiometric formula BiTl(S$_{1-\delta}$Se$_{\delta}$)$_2$,
    we mix a quantity $A$ computed in BiTlS$_2$ and BiTlSe$_2$ according to
  \begin{equation}
    A\big[ \text{BiTl(S}_{1-\delta}\text{Se}_{\delta}\text{)}_2 \big] =
      (1-\delta) A\big[\text{BiTlS}_2\big]
      + \delta A\big[\text{BiTlSe}_2\big]
    .
  \end{equation}
  The quantities $A$ being mixed are the dynamical matrix $\Phi$,
    the atomic masses $M$, the electron\textendash phonon squared coupling
    matrix $\Omega$ and the eigenvalues $\varepsilon$.
  In doing so, we keep track of the parity eigenvalue
    of the electronic states at $\Gamma$.
  The electronic quantities ($\Omega$, $\varepsilon$)
    are thus mixed between states with the same parity.

  In this work, we retain only the electron\textendash phonon coupling
    contribution to the self energy, and we neglect the effect
    of thermal expansion of the lattice.
  While the change of the volume as a function of doping is taken into account,
    the temperature dependence of the eigenvalues at a given doping
    is computed for a fixed-volume experiment.
  Our DFT and DFPT calculations
    \footnote{
    The ground state calculation is performed with an
    $8\times8\times8$ k-point grid and a kinetic energy cutoff
    of 50~Ha.
    The phonon wavevector sampling for the DFPT calculation is
    performed with an $8\times8\times8$ q-point grid for the
    full Brillouin zone, while the central region of the Brillouin zone
    is sampled with a $32\times32\times32$ q-point grid.
    }
    were performed with Abinit \cite{Gonze2009}
    using ONCV pseudopotentials \cite{hamann_optimized_2013}
    and a revised PBE functional \cite{zhang_comment_1998}.
  The choice of this exchange-correlation functional is motivated by
    the correct description of the ground state topology.
  An accurate description of the electronic structure would rely
    on $G_0W_0$ or self-consistent $GW$ calculations
    \cite{aguilera_spin-orbit_2013, aguilera_gw_2013}.
  However, the exchange-correlation functional used in this work
    yields the correct band topology for the materials under consideration,
    while allowing for the use of the DFPT method to obtain the lattice dynamics
    and the electron-phonon coupling.

  \subsection*{Results and discussion}

  \begin{figure}[tb]
    \includegraphics[width=0.8\linewidth]{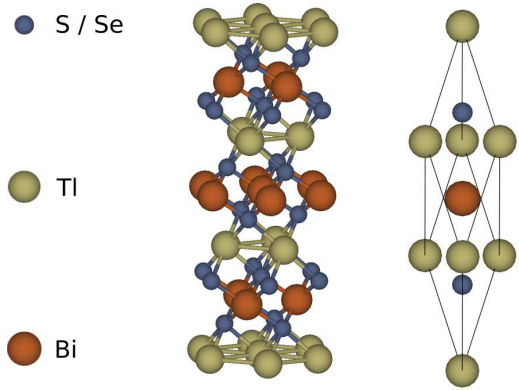}

    \caption{\label{fig:crystal}
      Crystal structure of BiTlS$_2$ or BiTlSe$_2$ showing
      the conventional unit cell (left) and the primitive unit cell (right).
      }
  \end{figure}

  The crystal structure of BiTlS$_2$ and BiTlSe$_2$
    is a close-packed stacking of hexagonal planes
    whose unit cell contains a single formula unit \cite{eremeev_textitab_2011},
    as shown in Fig.\ref{fig:crystal}.
  We obtained the lattice parameters by minimizing the internal stress
    ($<10^{-8}$Ha$/$Bohr$^3$),
    giving $a=4.207\AA$ and  $c=22.492\AA$ for BiTlS$_2$,
    and $a=4.372\AA$ and  $c=23.058\AA$ for BiTlSe$_2$,
    which are slightly overestimated compared to experiments
    ($a=4.1\AA$,  $c=21.9\AA$ for BiTlS$_2$,
    and $a=4.255\AA$,  $c=22.307\AA$ for BiTlSe$_2$) \cite{Xu2011a}.
  The only internal degree of freedom, $u$,
    is the relative height of the lowest sulfur or selenium atom.
  We relaxed the atomic coordinates until vanishing forces
    remained on the atoms ($<10^{-7}$Ha$/$Bohr),
    giving $u=0.237$ for BiTlS$_2$,
    and $u=0.239$ for BiTlSe$_2$.

  \begin{figure}[t]
    \includegraphics[width=\linewidth]{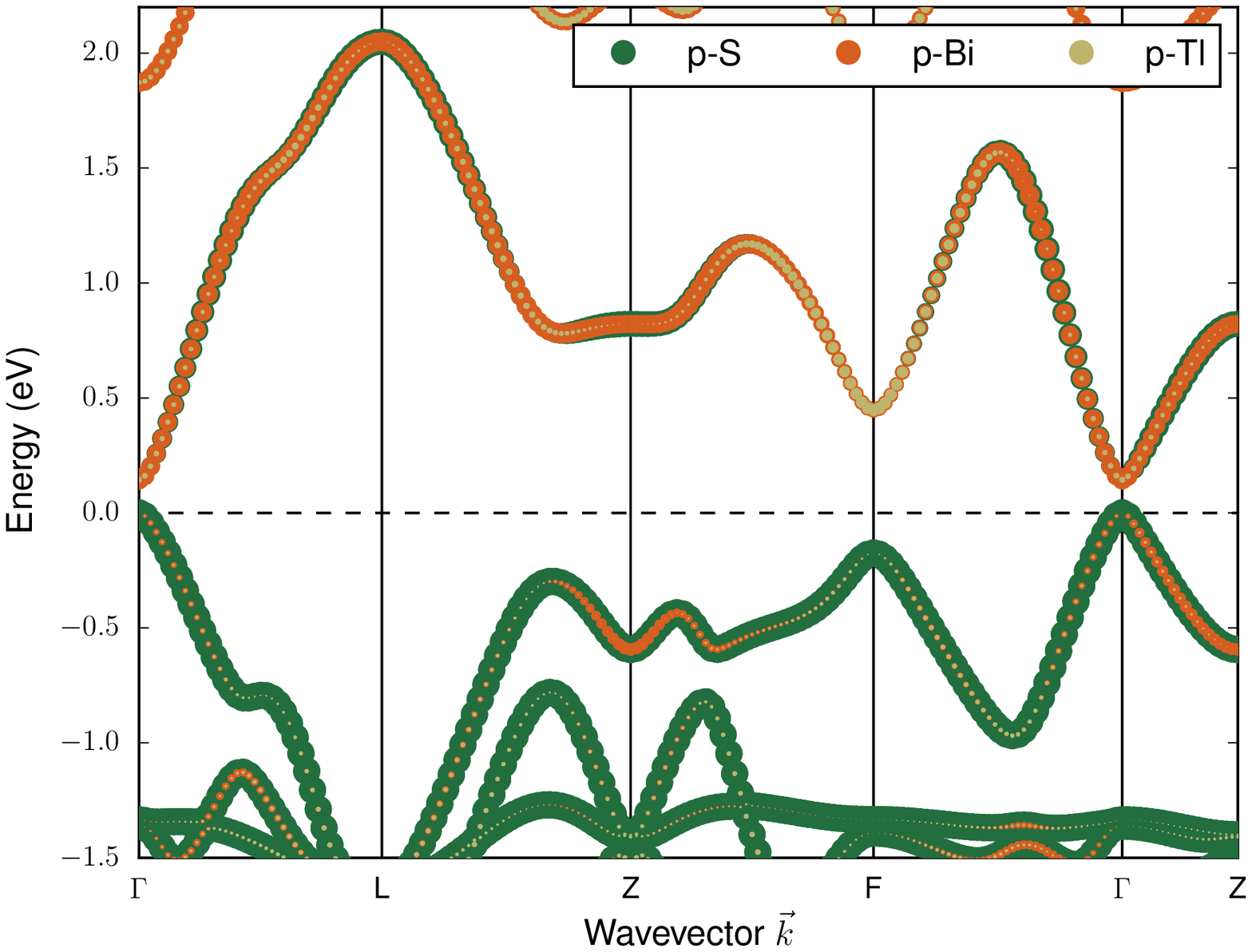}\\
    \includegraphics[width=\linewidth]{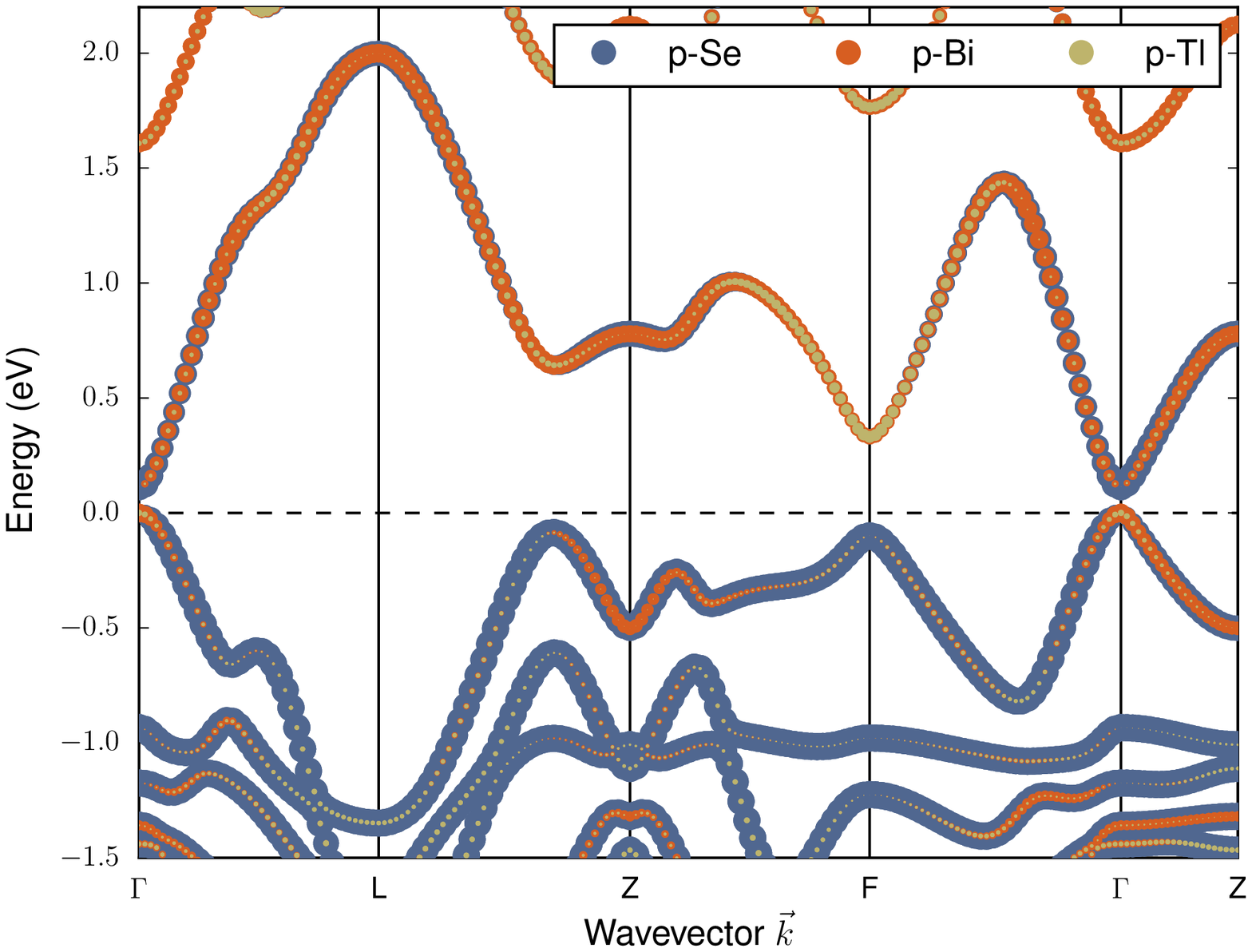}
    \caption{\label{fig:bandstructures}
      Band structures of BiTlS$_2$ (top) and BiTlSe$_2$ (bottom).
      The size of the colored discs is propotional to the projection
      of the electronic wavefunctions onto various angular momenta
      around the atoms.
      }
  \end{figure}

  The band structures of BiTlS$_2$ and BiTlSe$_2$ are quite similar
    in energy, but a distinct topology of the bands is revealed by the
    angular momentum decomposition of the electronic states,
    as shown in Fig.\ref{fig:bandstructures}.
  The $p$ states around thallium are always associated with a negative parity,
    since this atoms is an inversion center of the crystal
    and is taken as the origin in our calculations.
  The $p$ states around bismuth indicate a negative parity
    for the wavefunctions at $\Gamma$ and $F$,
    and a positive parity at $L$ and $Z$,
    since the application of inversion symmetry translates
    this atom into another primitive cell.
  In BiTlS$_2$, the characters of the last valence band and the first conduction
    band evolve smoothly through the Brillouin zone,
    resulting in a trivial phase with $Z_2=0$.
  In BiTlSe$_2$, 
    the characters of these bands invert at $\Gamma$,
    resulting in a topological phase with $Z_2=1$.

  \begin{figure*}[t]
    \includegraphics[width=\linewidth]{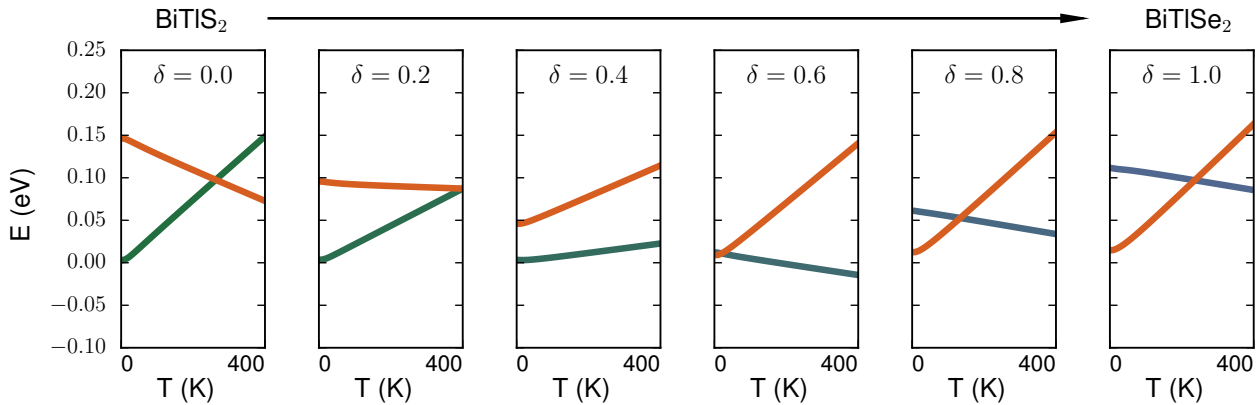}
    \caption{\label{fig:dopingtdep}
      Temperature dependence of the top of the valence bands and the bottom of
      the conduction bands for different doping between BiTlS$_2$ and BiTlSe$_2$.
      }
  \end{figure*}

  \begin{figure}[t]
    \includegraphics[width=\linewidth]{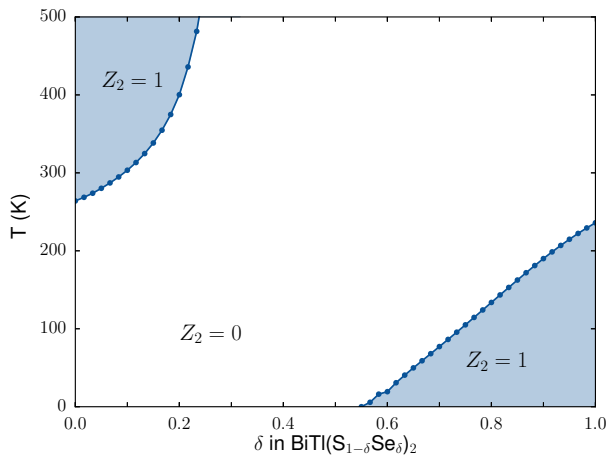}
    \caption{\label{fig:phasediagr}
      Topological phase diagram of BiTl(S$_{1-\delta}$Se$_{\delta}$)$_2$
      as a function of the doping parameter ($\delta$) and temperature.
      The blue shaded region indicates the topological phase.
      }
  \end{figure}

  Figure \ref{fig:dopingtdep}
    shows the temperature dependence of the valence bands maximum (VBM)
    and the conduction bands minimum (CBM)
    for various intermediate doping between BiTlS$_2$ and BiTlSe$_2$.
  By tracking the critical temperature as a function of doping,
    we obtain the corresponding topological phase diagram,
    shown in Fig.\ref{fig:phasediagr}.
  In BiTlS$_2$ and for low doping ($\delta \lesssim 0.3$),
    the band gap closes as a function of temperature,
    promoting the topological phase above the critical temperature.
  At intermediate doping,
    the electron\textendash phonon coupling self energy terms change sign
    but the system remains in the trivial state.
  The band gap now increases with temperature
    and no topological phase is found.
  Above the critical doping ($\delta \approx 0.55$) up to BiTlSe$_2$,
    the VBM and the CBM are inverted,
    and again the inverted band gap closes with temperature.
  The system is driven back into the trivial phase
    above the critical temperature.

  The sign flip of the self energy terms can be understood
    in terms of intra-band and inter-band scattering processes
    of the bands nearest to the band gap.
  These are the terms with the smallest energy denominators
    in the Fan self energy\textemdash
    the first term of Eq.\eqref{sigma}\textemdash
    making the dominant contributions to the eigenvalues renormalization.
  In the intra-band scattering, the VBM (CBM) couples 
    to another state in the same band with lower (higher)
    energy, and this process closes the band gap.
  Conversely, in the inter-band scattering, the VBM (CBM) couples 
    to a state in the first conduction band (last valence band),
    and this process opens the band gap.
  The strongest intra-band and inter-band interactions
    happen in the neighborhood of the $\Gamma$ and $F$ points in k-space,
    where the band gap reaches local minima.

  The relative strenght of inter-band and intra-band interactions
    stems from the symmetry of the coupling potential.
  Rewrite the electron\textendash phonon coupling elements as
    $g_{\vk n n'}(\vq,\ipho)=\bra{\vk+\vq n'} V_{\vq\ipho}(\vr) \ket{\vk n}$
    with the phonon potential
  \begin{equation}
    V_{\vq\ipho}(\vr) = 
      \sum_{\iat\icart} V_{\iat\icart} (\vq,\vr) \xi^{\ipho}_{\iat\icart}(\vq)
  .
  \end{equation}
  Due to inversion symmetry, the position $\bm{\tau}_{\iat}$ of an atom $\iat$
    is related to the position of its inversion partner $-\iat$
    in the same unit cell by
    $-\bm{\tau}_{\iat} = \bm{\tau}_{-\iat} + \mv I_\iat$,
  where $\mv I_\iat$ is a lattice vector.
  The consequence for the phonon polarization vectors is that
    inversion partners are related by
  \begin{equation}
    \xi^{\ipho}_{-\iat\icart}(\mv q) =
    \lambda_{\vq \ipho} e^{i\vq \cdot \mv I_{\iat}}
    \xi^{\ipho *}_{\iat\icart}(\mv q)
  ,
  \end{equation}
  with $\lambda_{\vq \ipho}=\pm 1$ defining the parity of the phonon vector.
  At time-reversal invariant momenta,
    the phonon potentials 
    are parity eigenfunctions
    with 
  \begin{equation}
    V_{\vq\ipho}(\vr) =-\lambda_{\vq \ipho}V_{\vq\ipho}(-\vr)
    .
  \end{equation}
  Therefore, a phonon with odd parity ($\lambda_{\vq \ipho}=+1$)
    can only couple electronic states with opposite parities,
    and a phonon with even parity ($\lambda_{\vq \ipho}=-1$)
    can only couple electronic states with the same parity.

  Since the parity of the bands is unchanged between $\Gamma$ and $F$,
    we can make the following statement about the phonon modes
    at these points.
  In both BiTlS$_2$ and BiTlSe$_2$,
    the even phonon modes will promote the topological phase,
    and the odd phonon modes will promote the trivial phase.
  Furthermore, we note that at $\Gamma$ and $F$,
    the even phonon modes are those where the pair of S or Se atoms move 
    in opposite directions, while the Bi and Tl atoms do not move.
  As the sulfur atoms are being substituted for the heavier selenium atoms,
    the coupling with even phonon modes decreases,
    and the odd phonon modes dominate.
  Therefore, the system transitions from a regime where the topological
    phase is promoted at high temperature
    to a regime where the trivial phase is promoted instead.

  In summary, we observed, from first-principles calculations,
    a temperature-induced band inversion occuring
    in BiTl(S$_{1-\delta}$Se$_{\delta}$)$_2$,
    and we computed the corresponding topological phase diagram
    in doping and temperature space.
  The non-trivial phase exists under two different regimes.
  In BiTlS$_2$ and for low doping,
    the topological phase is promoted above the critical temperature
    from a low-temperature trivial phase;
    in BiTlSe$_2$ and for high doping,
    the topological phase is observed only at low temperature
    and is suppresed above the critical temperature.
  Experimentally, non-trivial topological phases
    have been observed only at low temperatures so far.
  Our analysis indicates however that any topological insulator material
    containing light atoms forming inversion pairs
    could exhibit a topological phase that is promoted with temperature.

  \begin{acknowledgments}
  G. Antonius acknowledges fruitiful discussions with Ion Garate and Kush Saha.
  This work was supported by
    the National Science Foundation under Grant No. DMR15-1508412.
  The computational ressources were provided by the
    National Energy Research Scientific Computing Center (NERSC),
    a DOE Office of Science User Facility supported by the Office of Science
    of the U.S. Department of Energy
    under Contract No. DE- AC02-05CH11231,
    and the Extreme Science and Engineering Discovery Environment (XSEDE),
    which is supported by National Science Foundation
    Grant No. 787 ACI-1053575.
  \end{acknowledgments}

\end{document}